\documentclass[conference]{IEEEtran}

\usepackage{diagbox} 
\usepackage{graphicx}
\usepackage{epstopdf}
\usepackage{psfrag}
\usepackage{subfigure}
\usepackage{url}
\usepackage{stfloats}
\usepackage{amsfonts,amssymb,amsmath,bm,paralist,theorem,cite,ifthen,color}
\usepackage{caption}
\usepackage{calc}
\usepackage{enumerate}
\usepackage{array}
\usepackage{float}
\usepackage[ruled]{algorithm2e}

\def\BibTeX{{\rm B\kern-.05em{\sc i\kern-.025em b}\kern-.08em
    T\kern-.1667em\lower.7ex\hbox{E}\kern-.125emX}}

\newtheorem{proposition}{Proposition}
\newtheorem{remark}{Remark}

\newtheorem{corollary}{Corollary}

\newcommand{\epr}{\hfill\(\Box\)}

\graphicspath{{Figures/}}

\newcommand\Cc{\ensuremath{\mathcal{C}}}
\newcommand\Nc{\ensuremath{\mathcal{N}}}
\newcommand\Oc{\ensuremath{\mathcal{O}}}

\newcommand\Uc{\ensuremath{\mathcal{U}}}

\newcommand\Cs{\ensuremath{{\mathbb{C}}}}
\newcommand\Es{\ensuremath{{\mathbb{E}}}}

\newcommand\Zs{\ensuremath{{\mathbb{Z}}}}

\newcommand\Ab{\ensuremath{ \mathbf{A} }}

\newcommand\Db{\ensuremath{ \mathbf{D} }}
\newcommand\Fb{\ensuremath{ \mathbf{F} }}

\newcommand\Hb{\ensuremath{ \mathbf{H} }}
\newcommand\Ib{\ensuremath{ \mathbf{I} }}
\newcommand\Tb{\ensuremath{ \mathbf{T} }}
\newcommand\Pb{\ensuremath{ \mathbf{P} }}
\newcommand\Ub{\ensuremath{ \mathbf{U} }}

\newcommand\Vb{\ensuremath{ \mathbf{V} }}
\newcommand\Wb{\ensuremath{ \mathbf{W} }}
\newcommand\Jb{\ensuremath{ \mathbf{J} }}

\newcommand\Xb{\ensuremath{ \mathbf{X} }}

\newcommand\ab{\ensuremath{ \mathbf{a} }}
\newcommand\bb{\ensuremath{ \mathbf{b} }}
\newcommand\cb{\ensuremath{ \mathbf{c} }}
\newcommand\db{\ensuremath{ \mathbf{d} }}

\newcommand\nb{\ensuremath{ \mathbf{n} }}

\newcommand\ssb{\ensuremath{ \mathbf{s} }}

\newcommand\wb{\ensuremath{ \mathbf{w} }}

\newcommand\yb{\ensuremath{ \mathbf{y} }}

\newcommand\Gammab{\ensuremath{{\bm \Gamma}}}

\newcommand\Sigmab{\ensuremath{{\bm \Sigma}}}

\newcommand\diag{\ensuremath{{\rm diag}}}
\newcommand\tr{\ensuremath{{\rm Tr}}}

\IEEEoverridecommandlockouts

\begin{document}

\title{ Beam Squint Analysis and Mitigation via Hybrid Beamforming Design in THz Communications
}

\author{\IEEEauthorblockN{Mengyuan Ma, Nhan Thanh Nguyen and Markku Juntti}
\IEEEauthorblockA{Centre for Wireless Communications (CWC), Uninvesity of Oulu, P.O.Box 4500, FI-90014, Finland\\
Email: \{mengyuan.ma, nhan.nguyen, markku.juntti\}@oulu.fi}
}
%

\maketitle

\begin{abstract}
We investigate the beam squint effect in uniform planar arrays (UPAs) and propose an efficient hybrid beamforming (HBF) design to mitigate the beam squint in multiple-input multiple-output orthogonal frequency-division multiplexing (MIMO-OFDM) systems operating at terahertz band. We first analyze the array gain and derive the closed-form beam squint ratio that characterizes the severity of the beam squint effect on UPAs. The effect is shown to be more severe with a higher fractional bandwidth, while it can be significantly mitigated when the shape of a UPA approaches a square. We then focus on the HBF design that maximizes the system spectral efficiency. The design problem is challenging due to the frequency-flat nature and hardware constraints of the analog beamformer. We overcome the challenges by proposing an efficient decoupling design in which the digital and analog beamformers admit closed-form solutions, which facilitate practical implementations. Numerical results validate our analysis and show that the proposed HBF design is robust to beam squint, and thus, it outperforms the state-of-the-art methods in wideband massive MIMO systems.
\end{abstract}


%
\IEEEpeerreviewmaketitle

\section{Introduction}
Terahertz (THz) band communications with ultra-large bandwidth and antenna arrays are considered essential enablers for meeting the increasing demand for the data rate in the sixth generation (6G) wireless communications \cite{sarieddeen2021overview}. However, while the enormous bandwidth can increase the data rate, a myriad of antennas make fully digital beamforming practically infeasible. Therefore, hybrid beamforming (HBF) is often considered a low-complexity solution, which can achieve a good tradeoff between the spectral efficiency (SE) and energy efficiency (EE) for large or massive multiple-input multiple-output (MIMO) communications \cite{ahmed2018survey}.

The design of HBF in wideband systems is considerably challenging because the analog beamformer must be shared among the whole bandwidth. This motivates various HBF designs for MIMO orthogonal frequency-division multiplexing (MIMO-OFDM) systems \cite{yu2016alternating,sohrabi2017hybrid,tsai2018sub,zilli2021constrained,ma2021closed}. However, the frequency-selectivity of beamforming, known as {\it beam squint}, which can cause non-negligible performance loss, also needs to be addressed. To mitigate the beam squint effect, several approaches have been proposed  \cite{cai2016effect,liu2018space,dai2022delay,gao2021wideband,wu20223,ma2022switch,Nhan2022beam}. Specifically, wide beams can compensate for the loss in array gain induced by the beam squint effect \cite{cai2016effect,liu2018space}. Alternatively, different HBF structures have been explored to alleviate the beam squint. It has been shown in \cite{dai2022delay,gao2021wideband,wu20223} that high-resolution true-time-delayers (TTDs) embedded into the RF front-end can effectively mitigate the beam squint. However, the TTD at the THz band typically causes high power consumption, insertion loss, and hardware complexity \cite{yan2022energy}. Considering that the ultra-large numbers of antennas (e.g., thousands of antenna elements) could be deployed in THz transceivers, TTD-aided HBF with numerous TTDs can lead to an unacceptable encumbrance to the system EE. On the other hand, the switch-based HBF structures are more robust to the beam squint effect than the phase shifter-based HBF schemes, resulting in higher EE \cite{ma2022switch}. Furthermore, it has been recently observed in \cite{Nhan2022beam} that deploying uniform planar arrays (UPAs) rather than uniform linear arrays (ULAs) can significantly reduce the detrimental effect of beam squint. The above fact implies that the beam squint might be less of an issue if UPA is employed along with an efficient HBF design, which motivates us to investigate the potential of HBF design with UPA to overcome the beam squint effect in THz communication systems. 

We first analyze the array gain of a UPA-based system and derive a closed-form expression of beam squint ratio (BSR) that quantifies the severity of beam squint. A closed-form BSR has been obtained in \cite{ma2022switch} for ULA-based systems. 
We herein extend the investigation to the UPA and analytically justify that when the UPA is equipped with the same number of antennas in the horizontal and vertical dimensions, the beam squint effect is minimized. Specifically, with $N$ antennas and $\sqrt{N}$ being an integer, the UPA of size $\sqrt{N}\times \sqrt{N}$ can reduce the beam squint severity by $\sqrt{N}$ times compared to the ULA of size $N \times 1$.
Enlightened by the finding, we further design an efficient HBF scheme to mitigate the impact of the beam squint on the system SE. In the design, the formulated SE maximization problem is challenging due to the strong coupling of variables and hardware constraints of the analog beamformer. By decoupling the designs of precoders and combiners, we derive closed-form solutions to both. Numerical results validate our analysis and demonstrate that the proposed HBF scheme is more robust to the beam squint and capable of achieving higher SE compared to the state-of-the-art HBF designs for THz band communication systems.
 \vspace{5mm}
\section{THz MIMO-OFDM System}\label{sc:signal model}

We consider a single-user uplink MIMO-OFDM system where the mobile station (MS) and base station (BS) are equipped with $N_{\rm t}$ and $N_{\rm r}$ antennas, respectively. The MS sends the signal vector $\ssb[k] \in \Cs^{N_{\rm s} \times 1}$ of $N_{\rm s}$ data streams at the $k$th subcarrier, with $\Es\left[\ssb[k]\ssb^H[k]\right]=\Ib_{N_{\rm s}}, k=1,\cdots,K$, to the BS which deploys the fully-connected HBF architecture with $N_{\rm RF}$ RF chains ($N_{\rm s} \leq N_{\rm RF}\leq \min(N_{\rm t},N_{\rm r})$. We assume that a fully digital precoder $\Fb[k] \in \Cs^{N_{\rm t} \times N_{\rm s}}$ is employed at the MS, $ \| \Fb[k] \|^2_F \leq P[k]$, where $P[k]$ denotes the power budget for the $k$th subcarrier. At the BS, the received signal is first combined by $\Wb_{\rm RF}\in \Cs ^{N_{\rm r} \times N_{\rm RF}}$ in the analog domain and further processed in the baseband domain by $\Wb_{\rm BB}[k] \in \Cs ^{N_{\rm RF} \times N_{\rm s}}$ for subcarrier $k$. Note that $\Wb_{\rm RF}$ is the frequency-flat combining matrix with constant modulus entries. The post-processed signal at the $k$th subcarrier with the channel matrix $\Hb[k]\in \Cs^{N_{\rm r}\times N_{\rm t}}$ is expressed as
\begin{equation}\label{eq:received signal}
 \setlength{\abovedisplayskip}{5pt}
 \setlength{\belowdisplayskip}{5pt}
  \yb[k]=\Wb^H[k] \Hb[k] \Fb[k]\ssb[k] + \Wb^H[k]\nb[k],
\end{equation}
where $\Wb[k]=\Wb_{\rm BB}[k]\Wb_{\rm RF}$, $\nb_k\sim \Nc(\boldsymbol{0}, \sigma_{\rm n}^2\Ib_{N_{\rm r}})$ is the additive white Gaussian noise vector at the $k$th subcarrier with $\sigma_{\rm n}^2$ being the noise variance.

In the THz band communications, signal propagation experiences high attenuation and very limited scattering due to the short wavelength and molecular absorption. Since accurate modeling of the THz channel is of challenge, we adopt a statistical tap-delay profile modeling the impulse response and multipath parameters for ultra-broadband channels \cite{han2018propagation}. Assuming that UPA is utilized, the delay-$d$ channel response matrix for subcarrier frequency $f_{k}=f_{\rm c}+\left(k-\frac{K+1}{2} \right)\frac{B}{K}$ with central frequency $f_{\rm c} $ and bandwidth $B$ is given as \cite{yuan2020hybrid,tarboush2021teramimo}
\begin{equation}\label{eq:d delay channel}
 \setlength{\abovedisplayskip}{3pt}
 \setlength{\belowdisplayskip}{3pt}
 \small
   \Hb_d(f_k)=\zeta \sum\limits_{l=1}^{L_{\rm p}} \alpha_l p\left(d T_{\rm s}-\tau_l\right) \ab_{\rm r}\left( \theta_l^{\rm r},\phi_l^{\rm r}, f_k\right) \ab_{\rm t}^H\left( \theta_l^{\rm t},\phi_l^{\rm t}, f_k\right), 
\end{equation}
where $\zeta=\sqrt{\frac{N_{\rm r}N_{\rm t}}{L_{\rm p}}}$, $L_{\rm p}$ is the number of distinct scattering paths; $\alpha_l \sim \Cc\Nc(0,1)$ and $\tau_l$ are the complex channel gain and the delay of the $l$th path, respectively; $\theta_l^r$ ($\phi_l^{\rm r}$) and $\theta_l^t$ ($\phi_l^{\rm t}$) represent the azimuth (elevation) angles of arrival/departure (AoA/AoD) of the $l$th path, respectively; $T_{\rm s}$ is the sampling period; $p(\tau)$ denotes the pulse-shaping filter for $T_s$-spaced signaling evaluated at $\tau$ seconds \cite{alkhateeb2016frequency}.  Let $N_{\rm r,h} \times N_{\rm r,v}$ denote the size of the UPA at the BS, where $N_{\rm r,h} N_{\rm r,v} = N_{\rm r}$.  With $\rho_l^{\rm r}\triangleq  \sin(\theta_l^{\rm r})\sin(\phi_l^{\rm r})$ and $\varrho_l^{\rm r} \triangleq \cos(\theta_l^{\rm r}) $, the array response vector $\ab_{\rm r}\left( \theta_l^{\rm r},\phi_l^{\rm r}, f_k\right)$ at the BS can be expressed as\cite{li2020dynamic,yan2022energy}
\begin{equation}\label{eq:array steer vector for MIMO}
 \setlength{\abovedisplayskip}{4pt}
 \setlength{\belowdisplayskip}{4pt}
 \begin{aligned}
    \ab_{\rm r}\left( \rho_l^{\rm r},\varrho_l^{\rm r}, f_k\right) 
    =\ab_{\rm r,h}\left( \rho_l^{\rm r}, f_k\right) \otimes \ab_{\rm r,v}\left( \varrho_l^{\rm r}, f_k\right),
 \end{aligned}
\end{equation}
where $\otimes$ denotes the Kronecker product, and
\begin{equation}
 \setlength{\abovedisplayskip}{3pt}
 \setlength{\belowdisplayskip}{3pt}
 \small
\begin{aligned}
      &\ab_{\rm r,h}\left( \rho_l^{\rm r}, f_k\right)= \frac{1}{\sqrt{N_{\rm r,h}}}[1,e^{j2\pi\Delta_{\rm r,h}\frac{f_k}{f_{\rm c}}\rho_l^{\rm r}}, \cdots, 
      e^{j2(N_{\rm r,h}-1)\pi\Delta_{\rm r,h}\frac{f_k}{f_{\rm c}}\rho_l^{\rm r}}]^T,  \\
      &\ab_{\rm r,v}\left( \varrho_l^{\rm r}, f_k\right)= \frac{1}{\sqrt{N_{\rm r,v}}}[1,e^{j2\pi\Delta_{\rm r,v}\frac{f_k}{f_{\rm c}}\varrho_l^{\rm r}}, \cdots, 
 e^{j2(N_{\rm r,v}-1)\pi\Delta_{\rm r,v}\frac{f_k}{f_{\rm c}}\varrho_l^{\rm r}}]^T.
\end{aligned}
\end{equation}
Here, $\Delta_{\rm r,h} $ and $\Delta_{\rm r,v} $ denote the antenna spacing normalized by the wavelength of central frequency $f_{\rm c}$ at the horizontal and vertical dimension of the UPA, respectively. The array response vector $ \ab_{\rm t}\left( \theta_l^{\rm t},\phi_l^{\rm t}, f_k\right)$ at the MS can be modeled similarly. The frequency-domain channel at the $k$th subcarrier is given as $\Hb[k]=\sum_{d=0}^{D-1} \Hb_{d}(f_k) e^{-j\frac{2\pi k}{K}d}$ \cite{tse2005fundamentals}, where $D$ denotes the maximum channel delay taps.
\vspace{1mm}

\section{Beam Squint Effect in UPA}\label{sc:UPA beam squint}
Beam squint in a wideband multicarrier system with ULAs has been studied in the literature \cite{cai2016effect,dai2022delay,ma2022switch}. However, the results therein need to be adjusted for UPAs due to the different array structures. In the UPA, antennas deployed along two orthogonal dimensions provide more flexibility by allowing beamforming in all cardinal directions, which renders beam squint analysis more complicated. Focusing on the receive beamforming, we derive the closed-form BSR of a wideband system employing UPAs in Proposition \ref{prop:BSR closed form}. We note that even though the beam squint analysis herein is performed for the array at the receiver side, the results also apply to the transmitter with HBF architectures.
\begin{proposition}\label{prop:BSR closed form}
In wideband system employing an UPA of size $N_{\rm r,h} \times N_{\rm r,v}$, the BSR can be approximately given as
 \begin{equation}\label{eq:BSR closed-form}
  \setlength{\abovedisplayskip}{3pt}
 \setlength{\belowdisplayskip}{3pt}
   \text{ BSR} \approx \frac{b}{8}\max\left(N_{\rm r,h}\Delta_{\rm r,h}, N_{\rm r,v}\Delta_{\rm r,v}\right).
   \end{equation}
Here, $N_{\rm r,h}$ ($N_{\rm r,v}$) and $\Delta_{\rm r,h}$ ($\Delta_{\rm r,v}$) denote the number of antennas and the normalized antenna spacing of the UPA in horizontal (vertical) dimension, respectively, and $b \triangleq \frac{B}{f_{\rm c}}$ is the fractional bandwidth. 
\end{proposition}
{\it Proof:} The analog beamformer is typically designed based on the array response vector to maximize the array gain. In the HBF architecture, the $l$th column of the analog beamformer $\Wb_{\rm RF}$, i.e., the analog combing vector $\wb_{l}$ 
, can be used to generate the near-optimal beam towards the $l$th path's physical direction $(\theta_l^{\rm r},\phi_l^{\rm r})$ \cite{dai2022delay}. In this case, the normalized array gain at $k$th subcarrier is given as
\begin{equation}
 \setlength{\abovedisplayskip}{4pt}
 \setlength{\belowdisplayskip}{4pt}
    g\left(\wb_l,\theta_l^{\rm r},\phi_l^{\rm r}, f_k \right)=\left|\wb_l^H  \ab_{\rm r}\left( \theta_l^{\rm r},\phi_l^{\rm r}, f_k\right) \right|,
\end{equation}
where $|a|$ denotes the absolute value of the scalar $a$. For conventional analog beamforming, the combining vector $\wb_l$ is set to $\wb_l=\ab_{\rm r}\left( \theta_l^{\rm r},\phi_l^{\rm r}, f_c\right)$. Thus, we obtain
\begin{flalign}\label{eq:UPA array gain}
\small
       &g\left(\wb_l,\theta_l^{\rm r},\phi_l^{\rm r}, f_k \right)=\left|\ab_{\rm r}^H\left( \theta_l^{\rm r},\phi_l^{\rm r}, f_c\right)  \ab_{\rm r}\left( \theta_l^{\rm r},\phi_l^{\rm r}, f_k\right) \right| \nonumber  \\
        &=\left|\left( \ab_{\rm r,h}^H\left( \rho_l^{\rm r}, f_{\rm c}\right) \otimes \ab_{\rm r,v}^H\left(  \varrho_l^{\rm r}, f_{\rm c}\right) \right) \left( \ab_{\rm r,h}\left( \rho_l^{\rm r}, f_k\right) \otimes \ab_{\rm r,v}\left(  \varrho_l^{\rm r}, f_k\right) \right) \right| \nonumber \\
        &\overset{(\textrm{a})}{=} \left|\ab_{\rm r,h}^H\left( \rho_l^{\rm r}, f_{\rm c}\right)\ab_{\rm r,h}\left( \rho_l^{\rm r}, f_k\right) \right| \left|\ab_{\rm r,v}^H\left(  \varrho_l^{\rm r}, f_{\rm c}\right)\ab_{\rm r,v}\left(  \varrho_l^{\rm r}, f_k\right)  \right|,
\end{flalign}
where equality (a) follows the property $\left(\ab^H \otimes \bb^H\right) \left(\cb \otimes \db\right)= (\ab^H\cb) (\bb^H \db) $ with $\ab,\bb,\cb$ and $\db $ being vectors. By denoting $\xi_k \triangleq \frac{f_k}{f_{\rm c}} -1 $, we further have
\begin{align}\label{eq:NAG for ULA}
        &\left|\ab_{\rm r,h}^H\left( \rho_l^{\rm r}, f_{\rm c}\right)\ab_{\rm r,h}\left( \rho_l^{\rm r}, f_k\right) \right|=\frac{1}{N_{\rm r,h}}\left| \sum\limits_{n=0}^{N_{\rm r,h}-1} e^{j2n\pi \Delta_{\rm r,h}\left( \frac{f_k}{f_{\rm c}} -1\right)\rho_l^{\rm r}} \right| \nonumber \\
    &\overset{(\textrm{i})}{=}\left| \frac{\sin \left(N_{\rm r,h}\pi \Delta_{\rm r,h}\xi_k\rho_l^{\rm r} \right) }{N_{\rm r,h}\sin\left(\pi \Delta_{\rm r,h}\xi_k\rho_l^{\rm r}  \right)} e^{j (N_{\rm r,h}-1)\pi \Delta_{\rm r,h}\xi_k\rho_l^{\rm r} }\right| \nonumber \\
    &\overset{(\textrm{ii})}{=} \left| D_{N_{\rm r,h}}\left( \Delta_{\rm r,h}\xi_k\rho_l^{\rm r}\right) \right|,
\end{align}
where equality (i) follows from $\sum\limits_{n=0}^{N-1} e^{jnx}=\frac{\sin(Nx/2)}{\sin(x/2)}e^{j(N-1)x/2} $, and equality (ii), $D_N(x)\triangleq \frac{\sin(N\pi x)}{N\sin(\pi x)}$. With a similar expansion of the second absolute term in \eqref{eq:UPA array gain}, we obtain
\begin{equation}\label{eq:UPA array gain final}
 \setlength{\abovedisplayskip}{4pt}
 \setlength{\belowdisplayskip}{4pt}
 \small
     g\left(\wb_l,\theta_l^{\rm r},\phi_l^{\rm r}, f_k \right)
        =\left| D_{N_{\rm r,h}}\left( \Delta_{\rm r,h}\xi_k\rho_l^{\rm r}\right) \right| \left| D_{N_{\rm r,v}}\left( \Delta_{\rm r,v}\xi_k \varrho_l^{\rm r}\right) \right|.
\end{equation}

It can be observed from \eqref{eq:UPA array gain final} that $\wb_l$ can achieve the largest array gain for central frequency $f_c$, i.e., $g\left(\wb_l,\theta_l^{\rm r},\phi_l^{\rm r}, f_c \right) =1$, and approximately attain the largest array gain for other frequencies due to $f_k\approx f_c,\forall k$, i.e., $\xi_k\approx 0,\forall k$, in typical narrowband systems. However, this does not hold for wideband systems where $f_k$ is significantly different from $f_c$. Because the non-negligible physical direction deviations between $\frac{f_k}{f_{\rm c}}\rho_l^{\rm r}$ ($\frac{f_k}{f_{\rm c}}\varrho_l^{\rm r}$) and $ \rho_l^{\rm r}$ ($ \varrho_l^{\rm r}$), i.e., $\left|\xi_k\rho_l^{\rm r} \right| > 0$ ($\left|\xi_k \varrho_l^{\rm r} \right| > 0$), can cause significant array gain loss. Additionally, we have the following observations.
\begin{itemize}
\item  As shown in \eqref{eq:UPA array gain final}, the normalized array gain of a UPA can be factorized as the product of the gains of two ULAs along with the horizontal and vertical dimensions with parameters $(N_{\rm r,h}, \Delta_{\rm r,h})$ and $(N_{\rm r,v}, \Delta_{\rm r,v})$, respectively. Thus, array gain loss in either the horizontal or vertical ULAs can incur significant performance degradation.

    \item Note that the maximum normalized array gain $\left| D_N(x) \right| = 1$ is achieved at $x=0$ while $\left| x\right|\geq \frac{1}{N}$ moves outside the mainlobe of $\left| D_N(x) \right|$, wherein the normalized array gain is limited as $ \left| D_N(x) \right| \leq \frac{1}{N \sin\left(\frac{3\pi}{2N}\right)}$. Hence, we can conclude from \eqref{eq:NAG for ULA} that the loss of array gain is significant if $\left|\xi_k\rho_l^{\rm r} \right| \geq \frac{1}{N_{\rm r,h}\Delta_{\rm r,h}}$ for the horizontal ULA. Similar results can be obtained for the vertical ULA.
    
    
    \item  Since  $ \frac{1}{N_{\rm r,h}\Delta_{\rm r,h}}$ and $ \frac{1}{N_{\rm r,v}\Delta_{\rm r,v}}$ represent the half of the beamwidth in the horizontal and vertical ULAs, respectively \cite{tse2005fundamentals}, $\frac{\left|\xi_k \rho_l^{\rm r} \right|}{\frac{1}{N_{\rm r,h}\Delta_{\rm r,h}}} $ and $\frac{\left|\xi_k \varrho_l^{\rm r} \right|}{\frac{1}{N_{\rm r,v}\Delta_{\rm r,v}}}$ represents the relative offset between the squinted beam of subcarrier $k$ and the beam aligned with $f_{\rm c}$ along the horizontal and vertical dimension, respectively. 
\end{itemize}
Based on the above analysis, we define the BSR of a UPA as 
\begin{equation}\label{eq:BSR definition}
     \setlength{\abovedisplayskip}{4pt}
    \setlength{\belowdisplayskip}{4pt}
 \small
 \text{ BSR}\triangleq 
           \frac{1}{K} \sum\limits_{k=1}^K\max \left(\frac{1}{2}\int^{1}_{-1} \frac{\left|\xi_k \rho_l^{\rm r} \right|}{\frac{1}{N_{\rm r,h}\Delta_{\rm r,h}}} d\rho_l^{\rm r}, \frac{1}{2}\int^{1}_{-1} \frac{\left|\xi_k \varrho_l^{\rm r} \right|}{\frac{1}{N_{\rm r,v}\Delta_{\rm r,v}}} d\varrho_l^{\rm r}\right),
\end{equation}
which is the expectation (over all subcarrier frequencies and physical directions) of the maximum relative offsets along the vertical and horizontal dimensions of the UPA. It implies that the beam squint of a UPA is dominated by the most severe beam squint affecting one of its dimensions, i.e., the vertical and horizontal ones. 
The BSR in \eqref{eq:BSR definition} can be computed as
\begin{align}\label{eq:BSR closed-form}
  \text{ BSR}& =\frac{1}{2K}\sum\limits_{k=1}^K\left|\xi_k \right| \max \left(N_{\rm r,h}\Delta_{\rm r,h}, N_{\rm r,v}\Delta_{\rm r,v} \right) \nonumber\\
          &\overset{\rm (iii)}{\approx} \frac{b}{2}\int^{1}_{0}\left|x-\frac{1}{2} \right|dx\max\left(N_{\rm r,h}\Delta_{\rm r,h}, N_{\rm r,v}\Delta_{\rm r,v}\right) \nonumber\\
          &= \frac{b}{8}\max\left(N_{\rm r,h}\Delta_{\rm r,h}, N_{\rm r,v}\Delta_{\rm r,v}\right),  
\end{align}
where approximation ${\rm (iii)}$ follows from the facts that $K \gg 1$ and $\int^{1}_{0} f(x)dx=\displaystyle \lim_{K \to \infty} \frac{1}{K} \sum\limits_{k=1}^K f\left( \frac{k}{K} \right)$ for a continuous real-valued function $f(\cdot)$ defined on the closed interval $[0, 1]$ \cite{ma2022switch}. The derivation of \eqref{eq:BSR closed-form} completes the proof. \epr
  \begin{figure*}[htbp]
  \small
        \centering
        \hspace{-2mm}
        \subfigure[Normalized array gain of a UPA in horizontal and vertical dimensions.]
        { \label{fig:3D view} \includegraphics[width=0.34\textwidth]{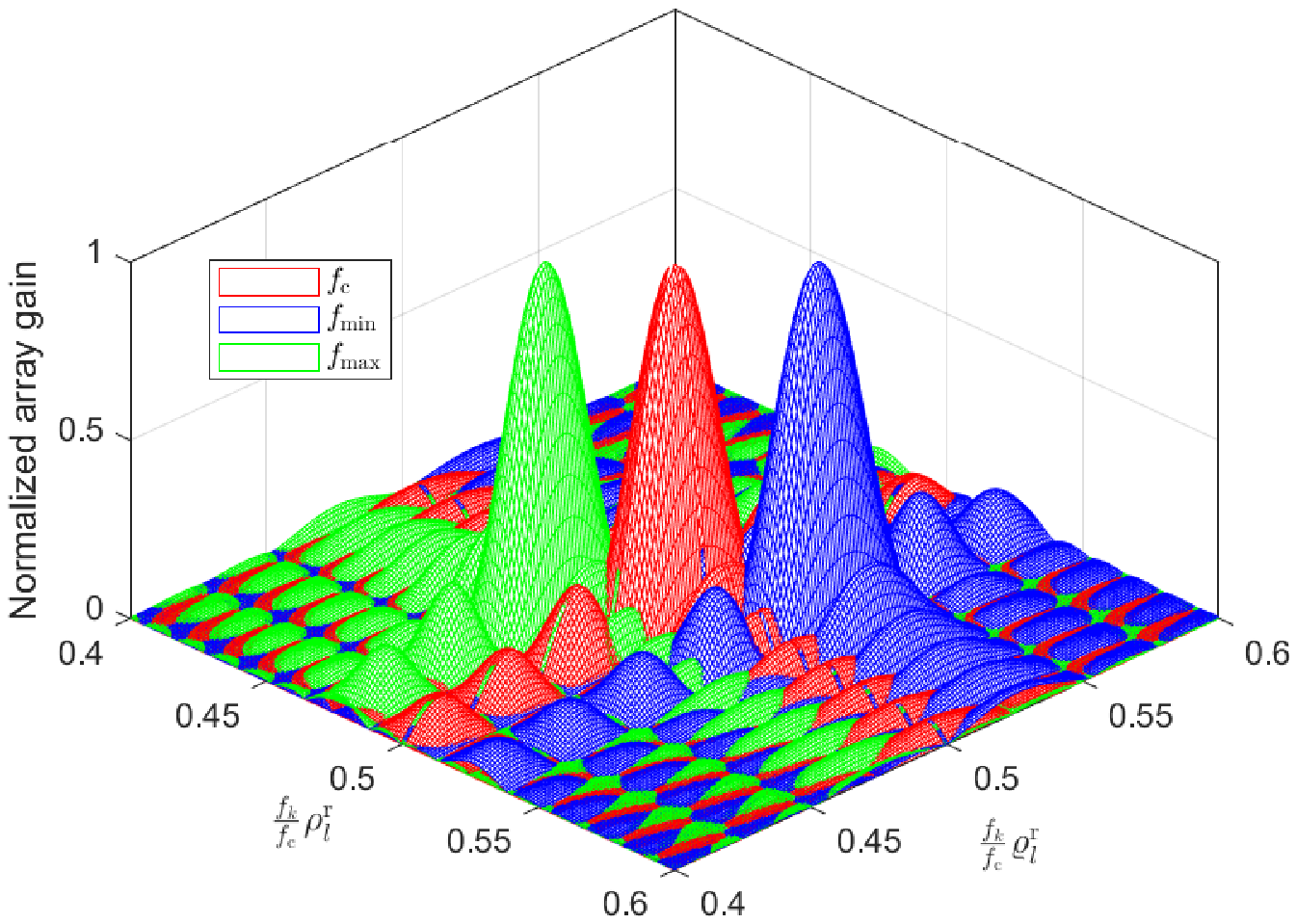}}
         \hspace{-7mm} 
        \subfigure[Horizontal view]
        {\label{fig:Front view}\includegraphics[width=0.34\textwidth]{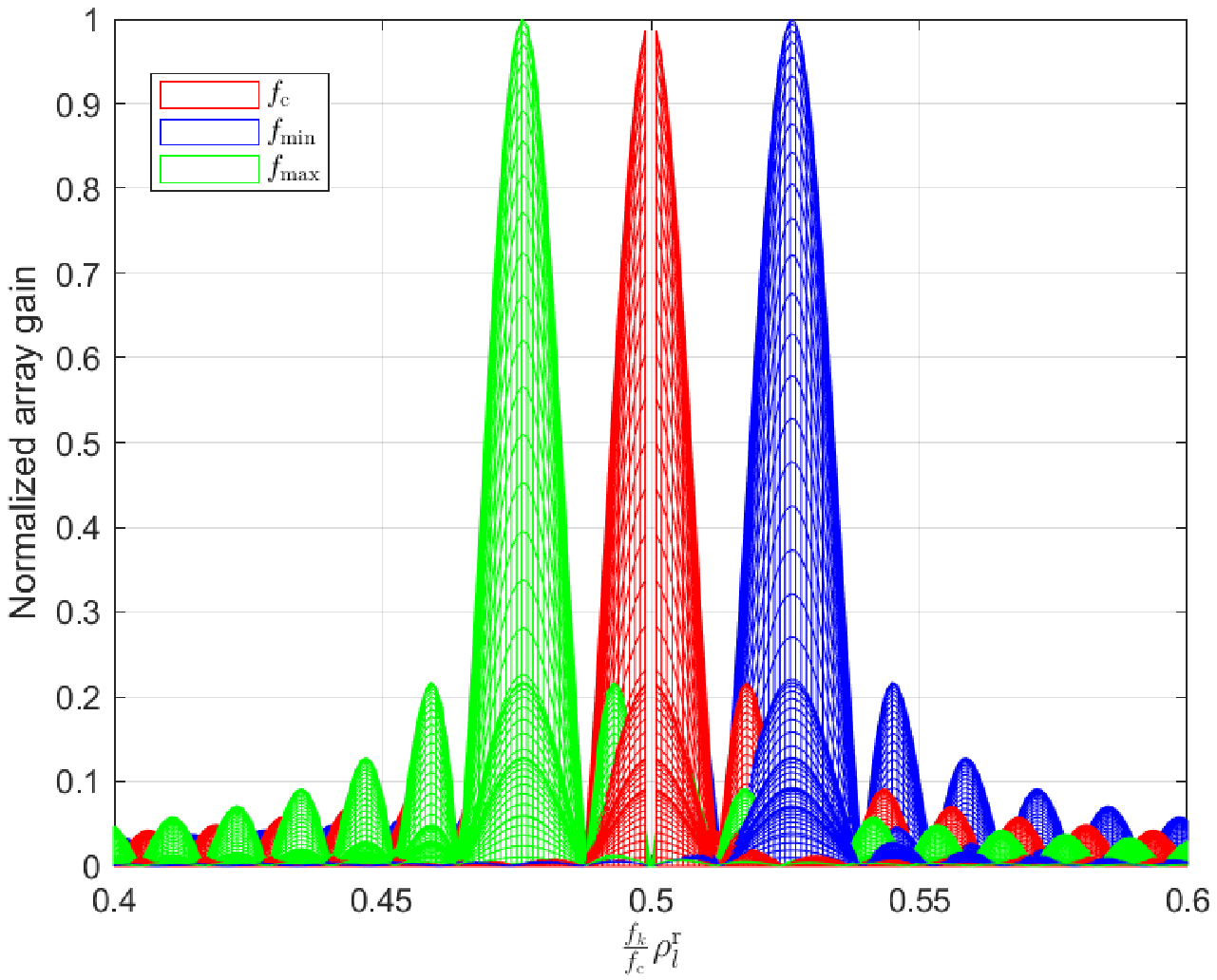}}
         \hspace{-8mm} 
        \subfigure[Vertical view.]
        {\label{fig:Profile view} \includegraphics[width=0.34\textwidth]{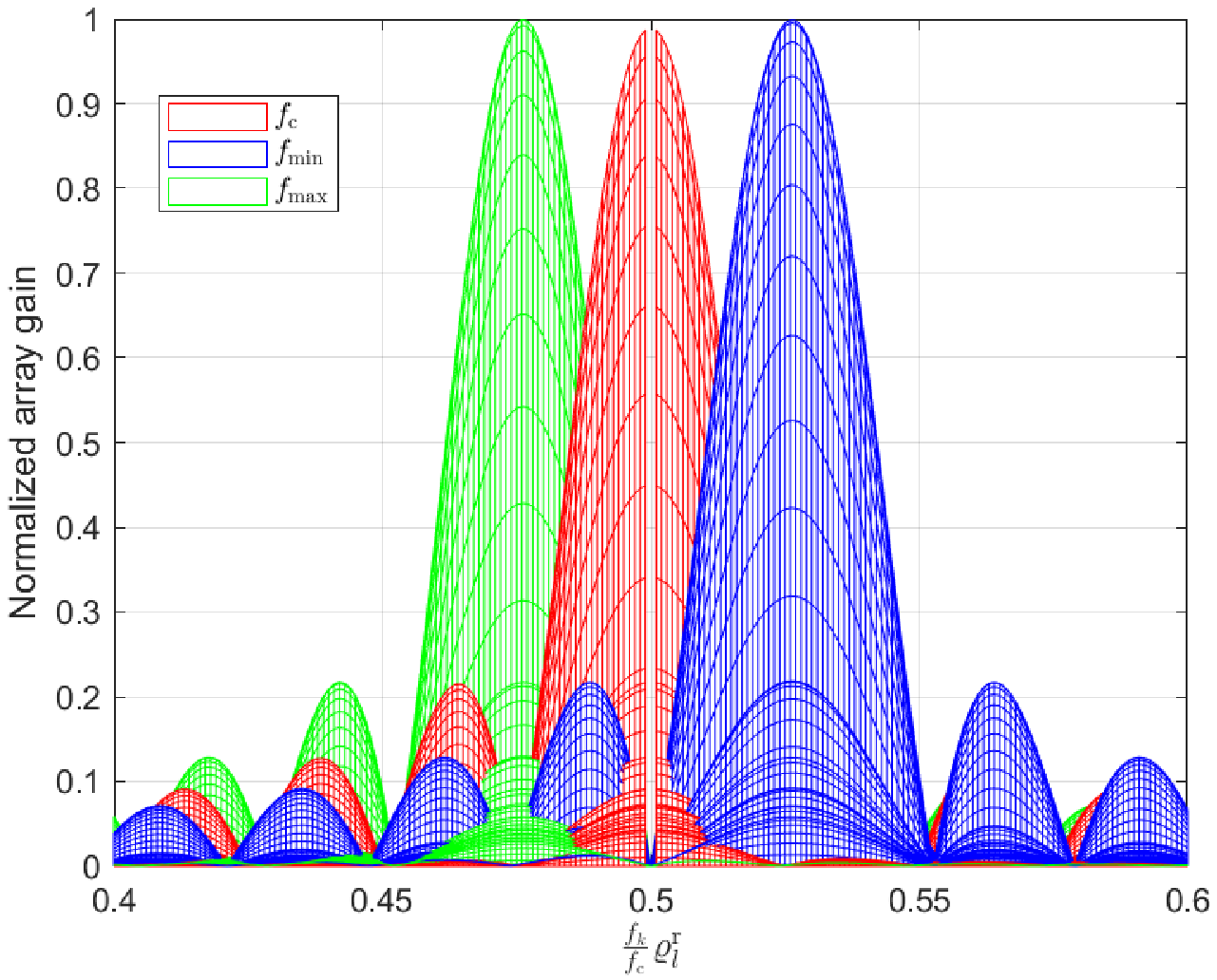}}
        \captionsetup{font={small}}
        \vspace{-0.3cm}
        \caption{ Normalized array gain achieved by $\wb_l$ versus $\left(\frac{f_k}{f_{\rm c}}\rho_l^{\rm r}, \frac{f_k}{f_{\rm c}}\varrho_l^{\rm r}\right)$ in systems with $f_{\rm c}=300~{\rm GHz},B=30~{\rm GHz}, K=128, \Delta_{\rm r,h}=\Delta_{\rm r,v}=\frac{1}{2}, N_{\rm r,h}=160, N_{\rm r,h}=80$, and $\left( \rho_l^{\rm r}, \varrho_l^{\rm r} \right)=\left(0.5, 0.5 \right)$, at the central frequency $f_{\rm c}$, the maximum frequency $f_{\rm max}$, and the minimum frequency $f_{\rm min}$.}
        \label{fig:Beam squint effect for UPA}
        	\vspace{-0.5cm}
  \end{figure*}
  
 We note that $\text{BSR}\geq 1$ implies that the squinted beam of subcarriers can be totally separate from the beam aligned with central frequency. This is because  $\text{BSR}\geq 1$ only arises when physical direction deviations of subcarrier $k$ exceed the half of beam mainlobe in either horizontal or vertical dimension, i.e., $\frac{\left|\xi_k \rho_l^{\rm r} \right|}{\frac{1}{N_{\rm r,h}\Delta_{\rm r,h}}} \geq 1$ or $\frac{\left|\xi_k \varrho_l^{\rm r} \right|}{\frac{1}{N_{\rm r,v}\Delta_{\rm r,v}}} \geq 1$.  As an example, we plot the normalized array gain of a wideband system with UPAs in Fig.\ \ref{fig:Beam squint effect for UPA} for $f_{\rm c}=300~{\rm GHz},B=30~{\rm GHz}, K=128, \Delta_{\rm r,h}=\Delta_{\rm r,v}=\frac{1}{2}, N_{\rm r,h}=160$, and $ N_{\rm r,v}=80$. According to \eqref{eq:BSR closed-form}, we obtain $\text{BSR}=\max(1, 0.5)=1$. This can be verified by results shown in Fig.\ \ref{fig:Front view} and Fig.\ \ref{fig:Profile view} where the physical derivations of the boundary frequencies, i.e., the maximum and minimum frequencies, along the horizontal and vertical dimensions exceed the beamwidth and the half of beamwidth of the mainlobe, respectively. Even though it is seen from the vertical view that the mainlobe of the beam at the central frequency can cover some parts of the boundary ones (see Fig.\ \ref{fig:Profile view}), these mainlobes are still completely separated, as occurred in the horizontal dimension (see in Fig.\ \ref{fig:Beam squint effect for UPA}(a)). Consequently, the frequency-flat analog beamformer cannot cover all the physical directions at all frequencies, causing severe performance loss. Moreover, based on Proposition \ref{prop:BSR closed form}, we obtain the optimal UPA configuration next.
\begin{corollary}\label{crlr: corollary}
Consider a UPA with half-wavelength antenna spacing, i.e., $\Delta_{\rm r,h}=\Delta_{\rm r,v}=\frac{1}{2}$, and $\sqrt{N_{\rm r}}\in \Zs_+$. Then, $N_{\rm r,h}=N_{\rm r,v}=\sqrt{N_{\rm r}}$ is the optimal configuration of the UPA that minimizes the BSR in \eqref{eq:BSR closed-form}, i.e., the square-shape UPA is the most robust to the beam squint effect. 
\end{corollary}
\begin{remark}
Based on Corollary \ref{crlr: corollary}, the minimum BSR for the systems employing UPAs is $\text{BSR}_{\rm upa}=\frac{\sqrt{N_{\rm r}}b}{16}$. Recalling that the BSR of ULAs is given as $\text{BSR}_{\rm ula}=\frac{N_{\rm r}b}{16}$ \cite{ma2022switch}, we have $ \frac{\text{BSR}_{\rm upa}}{\text{BSR}_{\rm ula}}=\frac{1}{\sqrt{N_{\rm r}}}$, which reveals that
 a UPA can reduce the beam squint effect by $\sqrt{N_{\rm r}}$ times compared to a ULA when both have the same number of antennas. Therefore, systems deploying UPAs rather than ULAs can significantly mitigate the beam squint effect, especially in THz communication systems with ultra-large arrays. 
\end{remark}

Furthermore, it was shown in \cite{ma2022switch} that the beam squint effect might be negligible for $\text{BSR}\leq 0.1$. Assume that the fractional bandwidth is $b=0.1$, which typically results to large bandwidth, e.g., $B=30~{\rm GHz}$ and $f_{\rm _c}=300~{\rm GHz}$. We further assume that the ULA or UPA adopts half-wavelength antenna spacing. According to the results in \cite{ma2022switch} for ULA, $N_{\rm r} \leq 16$ can lead to $\text{BSR}\leq 0.1$ while the UPA with up to $256$ antennas can achieve that. This is further verified by the numerical results in Section \ref{sc:simulation results}.

\section{HBF Design to Mitigate Beam Squint}
Motivated by results in Section \ref{sc:UPA beam squint}, in this section, we develop an efficient HBF scheme that can further enhance the system resistance to the beam squint effect. 

\subsection{Problem Formulation}
Based on \eqref{eq:received signal} and assuming the Gaussian signaling, the achievable SE of subcarrier $k$ can be expressed as
\begin{equation}\label{eq:SE for each subcarrier}
    \setlength{\abovedisplayskip}{4pt}
 \setlength{\belowdisplayskip}{4pt}
 \small
  R_k=\log_{2}\left|\Ib_{N_{s}}+\frac{1}{\sigma_n^{2}} \Wb[k]^{\dagger}  \Hb[k] \Fb[k] \Fb^H[k] \Hb^H[k] \Wb[k] \right|,
\end{equation}
where $\dagger$ denotes the Moore-Penrose inverse, and $\left|\Ab \right| $ represents the determinant of a matrix $\Ab$. Assuming that the 
the full channel state information (CSI) is available at both the MS and BS \cite{yu2016alternating,sohrabi2017hybrid,zilli2021constrained}, we aim at designing HBF matrices that maximize the overall SE of the system. The problem is formulated as
\begin{subequations}\label{eq:problem formulation}
  \begin{align}
    \underset{\Wb_{\rm RF},\atop \left\{\Fb[k], \Wb_{BB}[k]\right\}_{k=1}^{K}}{\max} \quad  & \frac{1}{K} \sum_{k=1}^{K} R_k \\
    \textrm{s.t.} \qquad  &  \| \Fb[k] \|^2_F \leq P[k] , \label{eq:precoder constraint}\\
    &\left|\Wb_{\rm RF}(i,j)\right|=1/\sqrt{N_{\rm r}},  \forall i, j, \label{eq: analog combiner constraint}
\end{align}
\end{subequations}
where $\Wb_{\rm RF}(i,j)$ denotes the entry of $\Wb_{\rm RF}$ in the $i$th row and the $j$th column. Problem \eqref{eq:problem formulation} is non-convex and of significant challenge to solve optimally. To tackle this, we decouple the designs of precoders $\Fb[k],\forall k$ and combiners $\Wb_{\rm RF},\Wb_{BB}[k], \forall k$, as elaborated next.

\subsection{Precoder Design}\label{sec:Transmitter precoder Design}
With given $\left\{ \Wb[k] \right\}_{k=1}^{K}$, we design the precoder as
  \begin{equation}\label{eq:problem formulation at transmitter}
    \setlength{\abovedisplayskip}{1pt}
 \setlength{\belowdisplayskip}{3pt}
        \max\limits_{ \left\{ \Fb[k]\right\}_{k=1}^{K} } \, \frac{1}{K} \sum_{k=1}^{K} R_k,\
        \text { subject to }  \eqref{eq:precoder constraint}, 
    \end{equation}
By defining $\Pb_t[k]=\Wb[k] \Wb[k]^{\dagger}$ and $\Hb_{\rm eff}[k]=\Pb_t[k]\Hb[k] $, $R_k$ can be rewritten as
\begin{equation}\label{eq:SE for transmitter}
 \setlength{\abovedisplayskip}{3pt}
 \setlength{\belowdisplayskip}{3pt}
  R_k
  = \log_{2}\left|\Ib+\frac{1}{\sigma_n^{2}} \Hb_{\rm eff}[k]\Fb[k] \Fb^H[k]  \Hb_{\rm eff}^H[k]  \right|.
\end{equation}
The optimal precoder $\Fb[k]$ for the $k$th subcarrier admits the water-filling solution as
\begin{equation}\label{eq:digital precoder}
 \setlength{\abovedisplayskip}{3pt}
 \setlength{\belowdisplayskip}{3pt}
    \Fb[k]=\Vb[k]\Sigmab^{\frac{1}{2}}[k],
\end{equation}
where $\Vb[k]$ is the matrix whose columns are the $N_{\rm s}$ right-singular vectors corresponding to $N_{\rm s}$ largest singular values $\left\{\lambda_1[k],\cdots, \lambda_{N_{\rm s}}[k]\right\}$ of $\Hb_{\rm eff}[k]$, and $\Sigmab[k]=\diag\left(p_1[k],\cdots,p_{N_{\rm s}[k]} \right)$ with $p_i[k]$ being the power allocated to the $i$th data stream at the $k$th subcarrier. Here, $ p_i[k]=\max\left(\mu-\frac{\sigma^2_{\rm n}}{\lambda_i[k]} \right)$, where $\mu$ is determined to meet $\sum\limits_{i=1}^{N_{\rm s}}p_i[k]= P[k]$.

\subsection{HBF Combiner Design}
For a fixed analog combiner $\Wb_{\rm RF}$, the optimal digital combiner of subcarrier $k$ can be obtained by the well-known MMSE solution \cite{sohrabi2017hybrid}
 \begin{equation}\label{eq:optimal baseband combiner}
     \setlength{\abovedisplayskip}{4pt}
 \setlength{\belowdisplayskip}{4pt}
   \Wb_{\rm BB}[k]=\left(\Jb[k]\Jb^H[k]+\sigma_{\rm n}^2\Wb_{\rm RF}^H\Wb_{\rm RF}\right)^{-1}\Jb[k],
 \end{equation}
 where $\Jb[k]\triangleq \Wb_{\rm RF}^H \Hb[k]\Fb[k]$. Here, we assume that $\Wb_{\rm BB}[k],\forall k $ and $\Wb_{\rm RF}$  are full-rank for the maximum spatial multiplexing gain. With $\Tb[k]\triangleq \Hb[k]\Fb[k]\Fb^H[k]\Hb^H[k]$, the SE of $k$th subcarrier is recast as $R_k = \log_2\left|\Ib+\frac{1}{\sigma_n^2}\Wb_{\rm RF}^{\dagger}\Tb[k]\Wb_{\rm RF}\right|$.
 Furthermore, the QR decomposition of $\Wb_{\rm RF}$ is given as $\Wb_{\rm RF}=\Ub_{\rm RF}\Db$, where $\Ub_{\rm RF}$ has orthogonal column vectors, i.e., $\Ub_{\rm RF}^H\Ub_{\rm RF}=\Ib_{N_{\rm RF}}$, and $\Db$ is an invertible matrix. With $\Wb_{\rm RF}=\Ub_{\rm RF}\Db$,  $R_k$ can be expressed as $R_k 
    =\log_2\left|\Ib+\frac{1}{\sigma_{\rm n}^2}\Ub_{\rm RF}^H\Tb[k]\Ub_{\rm RF}\right|$. Therefore, the problem of designing analog combiner $\Wb_{\rm RF} $ can be reformulated as
\begin{subequations}\label{eq:analog combiner design}
 \begin{align}
    \max\limits_{\Ub_{\rm RF}}  \quad & \; \frac{1}{K}\sum_{k=1}^{K}\log_2\left|\Ib+\frac{1}{\sigma_n^2}\Ub_{\rm RF}^H\Tb[k]\Ub_{\rm RF}\right| \label{eq:analog combiner design_a} \\
    \text { s.t. } \quad &\Ub_{\rm RF}^H\Ub_{\rm RF}=\Ib_{N_{\rm RF}}.
 \end{align}
 \end{subequations}
 Problem \eqref{eq:analog combiner design} is non-convex and challenging to solve optimally. Alternatively, we obtain an efficient suboptimal solution next. Let $\Tb[k]=\Xb[k]\Gammab[k]\Xb^H[k]$ be the truncated singular value decomposition (SVD) of $\Tb[k]$, where  $\Xb[k]\in \Cs^{N_{\rm r} \times N_{\rm s}}$ is a semi-unitary matrix and $\Gammab[k]\in \Cs^{N_{\rm s} \times N_{\rm s}}$, and let us define $\Tb_{\rm e}\triangleq \frac{1}{K}\sum_{k=1}^{K} \Xb[k]\Xb^H[k] $. The suboptimal solution to $\Ub_{\rm RF}$ can be obtained by solving the following problem
\begin{subequations}\label{eq:analog combiner desgin simple}
 \begin{align}
    \max\limits_{\Ub_{\rm RF}}  \quad & \; \tr\left(\Ub_{\rm RF}^H\Tb_{\rm e}\Ub_{\rm RF} \right) \\
    \text { s.t. } \quad &\Ub_{\rm RF}^H\Ub_{\rm RF}=\Ib_{N_{\rm RF}},
 \end{align}
\end{subequations}
which admits optimal solution $\Ub_{\rm RF}^* = \Ub_{\rm e}$, where the columns of $\Ub_{\rm e} $ are the $N_{\rm RF}$ eigenvectors associated with the $N_{\rm RF}$ largest eigenvalues of $\Tb_{\rm e}$ \cite{lutkepohl1997handbook}. In \eqref{eq:analog combiner desgin simple}, the objective is obtained by a Taylor expansion of \eqref{eq:analog combiner design_a}. We omit the detailed proof here due to limited space. 
 
 Finally, the solution to $\Wb_{\rm RF}$ can be derived by projecting the solution $\Ub_{\rm RF}^*$ of problem \eqref{eq:analog combiner design} into the feasible space of analog beamformer \cite{ma2021closed}, i.e.,
\begin{equation}\label{eq:optimal solution to analog combiner}
    \setlength{\abovedisplayskip}{3pt}
 \setlength{\belowdisplayskip}{3pt}
  \Wb_{\rm RF}(i,j)=\frac{1}{\sqrt{N_{\rm r}}}e^{j\angle\Ub_{\rm RF}^*(i,j)},\; \forall i,j,
\end{equation}
where $\angle x$ denotes the phase of a complex number $x$. We note that because the frequency-flat analog combiner is obtained based on the eigenvectors related to all subcarriers rather than to only the carrier frequency, it can better adapt to the common characteristics of subcarriers, making it less affected by the beam squint on average. This will be validated in Section \ref{sc:simulation results}.   
 
 The proposed HBF algorithm is summarized in Algorithm \ref{alg:HBF algorithm}, where the $\Wb[k]$ is initialized as the optimal digital combiner, i.e., $\Wb[k]=\Ub[k]$, with $\Ub[k]$ being a semi-unitary matrix whose columns are the $N_{\rm s}$ left singular vectors associated with the largest $N_{\rm s}$ singular values of $\Hb[k]$. Considering $N_{\rm t}\ll N_{\rm r} $, the overall computational complexity of Algorithm \ref{alg:HBF algorithm} is $\Oc\left( 2KN_{\rm t}N_{\rm r}^2\right) $, which is mainly caused by performing SVD and matrix multiplications.
 \vspace{-0.3cm}
\begin{algorithm}[hbpt]
\small
\caption{HBF Design for Problem \eqref{eq:problem formulation}}\label{alg:HBF algorithm}
\LinesNumbered 
\KwIn{$\Hb[k],\Wb[k],P[k], \forall k$, $\sigma^2_{\rm n}$}
\KwOut{$\Wb_{\rm RF} $, $\Fb[k],\Wb_{\rm BB}[k], \forall k$}
Obtain $\Fb[k]$ based on \eqref{eq:digital precoder} with $\small{\Hb_{\rm eff}[k]=\Wb[k] \Wb[k]^{\dagger}\Hb[k]}$.

Perform SVD $\Hb[k]\Fb[k]\Fb^H[k]\Hb^H[k]=\Xb[k]\Gammab[k]\Xb^H[k],\forall k$.

Obtain $\Wb_{\rm RF} $ as in \eqref{eq:optimal solution to analog combiner} with $\Tb_{\rm e}=\frac{1}{K}\sum_{k=1}^{K} \Xb[k]\Xb[k]^H $.

Obtain $\Wb_{\rm BB}[k]$ according to \eqref{eq:optimal baseband combiner}.
\end{algorithm}
\vspace{-0.4cm}
\section{Simulation Results}\label{sc:simulation results}
We herein present numerical results to verify our analysis and evaluate the performance of the proposed HBF scheme. In the simulations, we set $N_{\rm RF}=N_{\rm s}=4 $ at the BS and $N_{\rm t,h}=N_{\rm t,v}= 4$ and half-wavelength spacing in the UPA at the MS. The parameters of the channel model in \eqref{eq:d delay channel} are set to $L_{\rm p}=4,K=128,f_{\rm c}=300~{\rm GHz} $, $\theta_l^{\rm r}, \theta_l^{\rm t}\sim \Uc\left(-\pi, \pi \right) $, $\phi_l^{\rm r},\phi_l^{\rm t}\sim \Uc\left(-\frac{\pi}{2}, \frac{\pi}{2} \right)$, and the pulse shaping filter is modeled by the raised cosine function \cite{alkhateeb2016frequency} with the roll-off factor being $1$.
 The path delay is uniformly distributed in $[0,(D-1)T_{\rm s}]$ where $D$ is the cyclic prefix length, given by $D=K/4$ according to the specification 802.11ad  \cite{alkhateeb2016frequency}. The SNR is defined as SNR$~\triangleq \frac{P_{\rm b}}{\sigma_{\rm n}^2}$ where $P_{\rm b}= P[k],\forall k$. The other parameters are detailed in each figure. All reported results are averaged over $10^3$ channel realizations. For comparison, we consider the HBF-LSAA algorithm in \cite{sohrabi2017hybrid}, and the HBF-CFUBM algorithm in \cite{ma2021closed}, which are designed by maximizing the SE upper bound. In addition, the HBF-DCF algorithm in \cite{Nhan2022beam} designed for the central frequency, as is the typical method of designing analog beamformers in narrowband systems, is also included. The achievable SE of optimal digital beamforming (DBF) via the water-filling algorithm is denoted as $\text{R}_{\rm opt}$.

\begin{figure}[htbp]
  \vspace{-0.5cm}
        \centering
        \subfigure[$N_{\rm r}=256, \text{SNR}=10~{\rm dB}$.]
        {\label{fig:RSI}\includegraphics[scale=0.53]{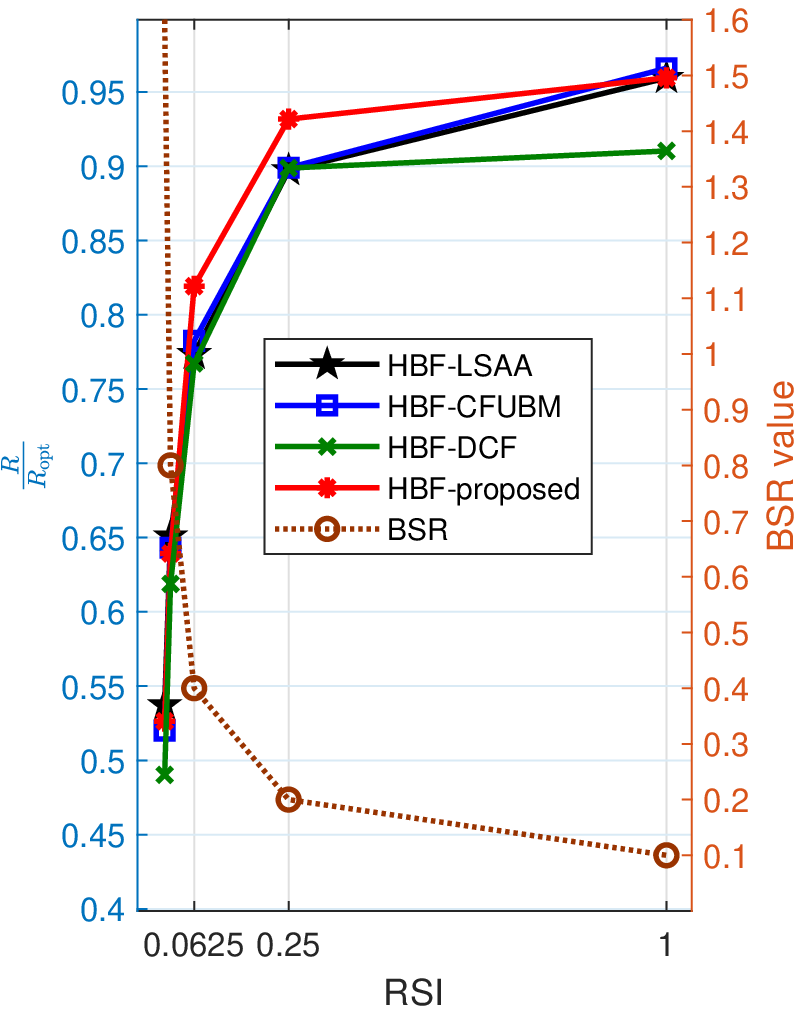}}
        \subfigure[ $N_{\rm r,h}=N_{\rm r,v}= 64$. ]
        {\label{fig:SNR} \includegraphics[scale=0.53]{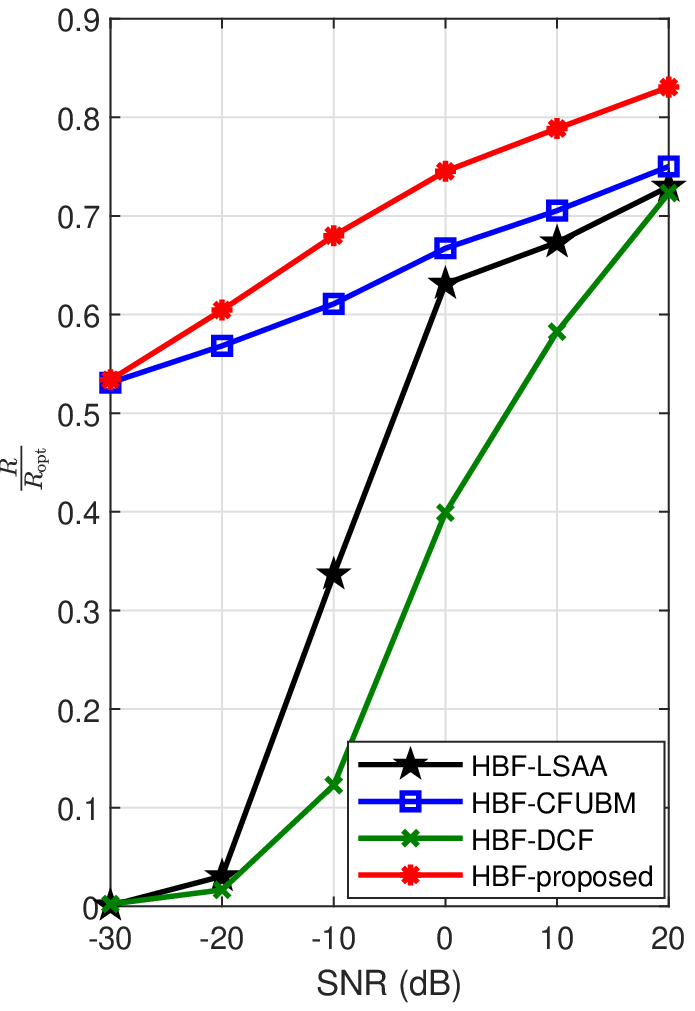}}
        \captionsetup{font={small}}
        \vspace{-0.3cm}
        \caption{ Normalized achievable SE versus RSI and SNR with $B=30~{\rm GHz}$ and $\Delta_{\rm r,h}=\Delta_{\rm r,v}=\frac{1}{2}$.}
        \label{fig:performance vs RSI and SNR}
        	\vspace{-0.3cm}
  \end{figure}

To verify Corollary \ref{crlr: corollary}, we consider the regular shape indicator (RSI) of a UPA as the ratio between the UPA's width $W_{\rm upa}$ and length $L_{\rm upa}$, i.e., $\text{RSI}\triangleq\frac{W_{\rm upa}}{L_{\rm upa}} \in (0,1]$. $\text{RSI}_{\rm upa} \rightarrow 1$ implies the UPA shape is close to a square, while $\text{RSI}_{\rm upa} \rightarrow 0$ indicates that the UPA is close to a ULA. Fig.\ \ref{fig:RSI} shows the performance of considered HBF algorithms versus RSI with $N_{\rm r}=256, B=30~{\rm GHz}, \text{SNR}=10~{\rm dB}$ and $\Delta_{\rm r,h}=\Delta_{\rm r,v}=\frac{1}{2}$. Note that the total number of receive antennas is fixed, while $N_{\rm r,h}\in \left\{1, 2, 4, 8, 16\right\}$ and $N_{\rm r,v}=\frac{N_{\rm r}}{N_{\rm r,h}}$. In this figure, the achievable SE of HBF algorithms are normalized by $\text{R}_{\rm opt}$. It can be observed from Fig.\ \ref{fig:RSI} that the ULA causes high BSR, i.e., severe beam squint effect, which significantly degrades the performance of HBF algorithms. Specifically, those HBF algorithms only attain about 50\% of the optimal achievable SE. When the RSI increases, i.e., the UPA gradually approaches a square shape, the BSR considerably decreases to its minimum, thereby improving the achievable SE of HBF algorithms. Note that with $N_{\rm r}= 256$ ($\text{BSR}= 0.1$), the considered HBF algorithms, except for the HBF-DCF scheme, can attain up to 95\% of the optimal performance. It is also shown that the proposed HBF algorithm is more robust to the beam squint effect compared to other HBF algorithms, which will be further justified in the following results.
  
Fig.\ \ref{fig:SNR} shows the normalized achievable SE as a function of SNR with $N_{\rm r,h}=N_{\rm r,v}= 64, B=30~{\rm GHz}$ and $\Delta_{\rm r,h}=\Delta_{\rm r,v}=\frac{1}{2}$, and thus, $\text{BSR}=0.4 $ based on \eqref{eq:BSR closed-form}. It is observed that the proposed HBF scheme achieves the best performance, and its superior performance is more clearly seen for low SNRs. Specifically, the proposed HBF scheme can attain over 65\% of the optimal achievable SE even at SNR $=-10$ dB, which is significantly higher than those of HBF-LSAA and HBF-DCF and 7\% higher than that of the HBF-CFUBM. This is because we design the analog beamformer based on the eigenvectors related to all subcarriers, while the other three HBF schemes do not. Furthermore, at high SNRs, e.g., at SNR $=20$ dB, the proposed HBF scheme can achieve over 80\% of the optimal performance while those of the others are under 75\%. 
    \begin{figure}[htbp]
    \vspace{-0.1cm}
        \centering
        \subfigure[$\Delta_{\rm r,h}=\Delta_{\rm r,v}=\frac{1}{2}$.]
        {\label{fig:SE vs bandwidth half-spacing}\includegraphics[scale=0.55]{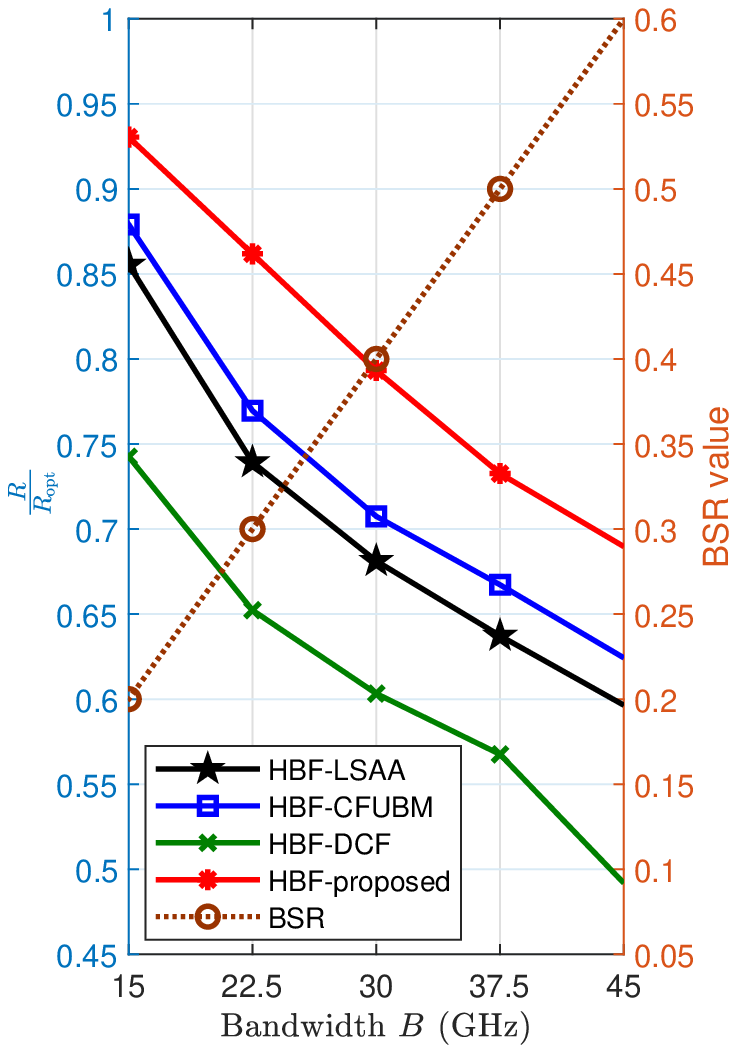}}
        \subfigure[ $\Delta_{\rm r,h}=\Delta_{\rm r,v}=\frac{1}{4}$. ]
        {\label{fig:SE vs bandwidth quarter-spacing} \includegraphics[scale=0.55]{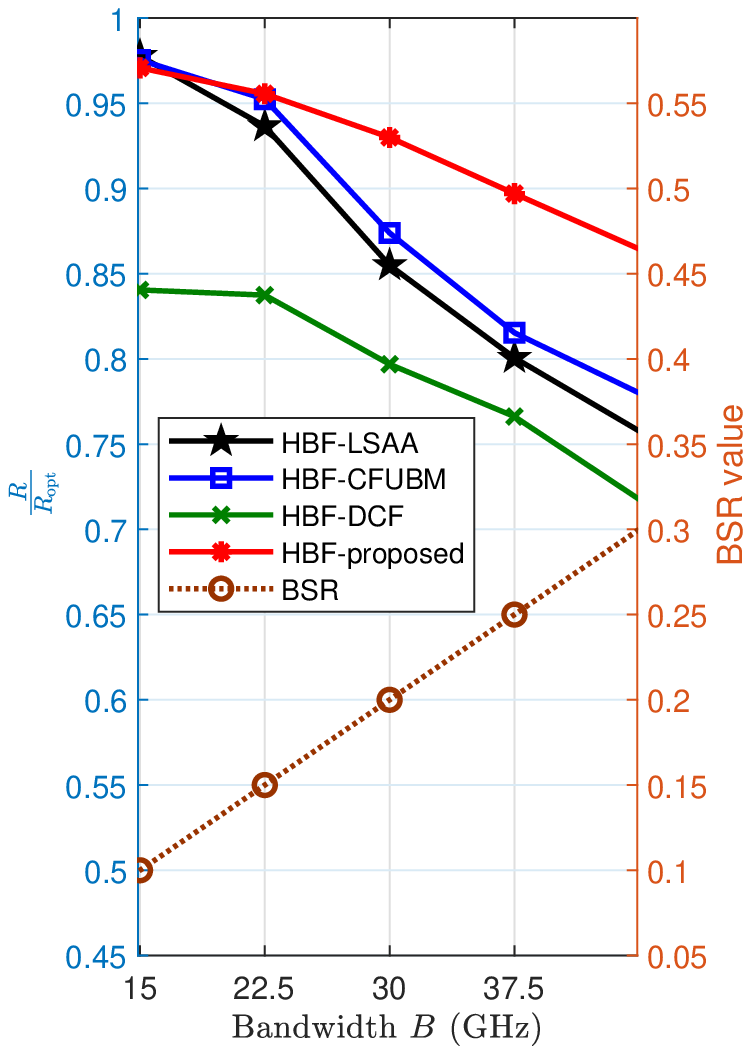}}
        \captionsetup{font={small}}
        \vspace{-0.2cm}
        \caption{Normalized achievable SE versus bandwidth with $N_{\rm r,h}=N_{\rm r,v}= 64, \text{SNR}=10~{\rm dB}$.}
        	\label{fig:Bandwidth performance}
        	\vspace{-0.7cm}
  \end{figure} 
  
The performance of the considered HBF algorithms with respect to bandwidth is presented in Fig.\ \ref{fig:Bandwidth performance} with $N_{\rm r,h}=N_{\rm r,v}= 64$, $ \text{SNR}=10~{\rm dB}$, and $\Delta_{\rm r,h}=\Delta_{\rm r,v}=\left\{\frac{1}{2},\frac{1}{4} \right\}$. We can observe from Fig.\ \ref{fig:SE vs bandwidth half-spacing} that the BSR linearly increases with the bandwidth, and consequently, the performance of HBF algorithms degrades significantly from 96\% to below 70\% of the optimal one. However, the achievable SE of those HBF schemes is considerably enhanced when the antenna spacing decreases, as shown in Fig.\ \ref{fig:SE vs bandwidth quarter-spacing}. This is because the severity of the beam squint effect is reduced by half when the antenna spacing is halved as indicated in Proposition \ref{prop:BSR closed form}. As a result, the performance of the HBF schemes is significantly improved, especially for large bandwidth. For example, with $B=45$ GHz, the proposed HBF scheme achieves the normalized SE of $0.86$ with $\Delta_{\rm r,h}=\Delta_{\rm r,v}=\frac{1}{4}$, which is nearly $17\%$ improvement compared to the case of $\Delta_{\rm r,h}=\Delta_{\rm r,v}=\frac{1}{2}$. Moreover, it is seen that the proposed HBF algorithm is more robust to the beam squint effect than other compared HBF schemes and capable of achieving a higher SE than HBF-LSAA and HBF-CFUBM for $\text{BSR}\geq 0.15$, as shown in Fig.\ \ref{fig:SE vs bandwidth quarter-spacing}. In addition, the performance of the considered HBF algorithms, except for the HBF-DCF scheme, can achieve over 96\% of the optimal achievable SE for $\text{BSR}\leq 0.1$, which coincides with that in \cite{ma2022switch}. 

\section{Conclusion}
This paper investigates the beam squint effect of wideband systems deploying UPA and proposes an efficient HBF design less affected by beam squint in THz communications. We first derive the closed-form BSR and analytically show that the beam squint effect is mitigated most when the numbers of antennas in the horizontal and vertical dimensions of the UPA approach the same value. Particularly, systems deploying square-shaped UPAs can significantly reduce the beam squint compared to that employing ULAs, especially in large-array systems. We then propose an efficient HBF design that maximizes the SE. Closed-form solutions for both digital precoders and HBF combiners are obtained. The numerical results validate our analysis and demonstrate that the proposed HBF algorithm is robust to the beam squint and, thereby capable of outperforming state-of-the-art HBF schemes.

\bibliographystyle{IEEEtran}
\bibliography{conf_short,jour_short,SW_HBF}

\end{document}